\documentclass[12pt]{article}
\usepackage{amsfonts,amssymb,amsmath,latexsym}
\newcommand{\be}{\begin{equation}}
\newcommand{\ee}{\end{equation}}
\newcommand{\bea}{\begin{eqnarray}}
\newcommand{\eea}{\end{eqnarray}}
\newcommand{\ba}{\begin{array}}
\newcommand{\ea}{\end{array}}
\newcommand{\bt}{\begin{tabular}}
\newcommand{\et}{\end{tabular}}

\newcommand{\fr}{\frac}
\newcommand{\ci}{\cite}
\newcommand{\cl}{\centerline}
\newcommand{\bs}{\bigskip}
\newcommand{\sms}{\smallskip}
\newcommand{\vs}{\vspace}

\newcommand{\en}{\eqno}

\newcommand{\bbib}{\begin{thebibliography}}
\newcommand{\ebib}{\end{thebibliography}}

\newcommand{\mbb}{\mathbb}

\newcommand{\und}{\underline}
\newcommand{\np}{\newpage}

\begin{document}

\titlepage
\vs{2cm}

\cl{\bf D-DIMENSIONAL CONFORMAL  $\sigma$-MODELS}
\cl{\bf AND THEIR TOPOLOGICAL EXCITATIONS }
\vs{0.5cm}
\cl{\bf S.A.Bulgadaev}
\vspace{0.5cm}
\centerline{L.D.Landau Institute for Theoretical Physics}
\centerline{Moscow, RUSSIA}
\centerline{Max Plank Institute for Physics of Complex Systems}
\centerline{Dresden, GERMANY}

\vs{0.5cm}
\centerline{A talk presented at the International Conference ICMP-2000,}
\cl{17 - 22 July 2000, Imperial College, London, England}
\vs{0.5cm}
\np

\cl{\bf I. INTRODUCTION}
\bs

The nonlinear $\sigma$-models (NSM) play an important role in
the modern theoretical physics (see, for example \ci{1,2}).
Usually, being an effective local theories,
they describe the low-energy and long-wave behaviour of systems
with degenerate vacuum manifolds ${\cal M}$.

The most interesting items in NSM are
the existence and structure of the gapless modes spectra.
The answers on these questions have some universality and depend only on
dimension of space, symmetry and topology  of ${\cal M}.$

The most important are the two-dimensional NSM, which have many
interesting properties, including, at the classical level,

1) locality,

2) scale invariance,

and, on the quantum level,

3) renormalizability.

In some cases they have the next additional properties:

4) scale symmetry breaking and  mass generation \ci{1},

5) topological excitations (TE),

6) integrability (both on classical and quantum levels \ci{3}).

The passage to other dimensions brings a lost of some of these properties.
If, for example, one tries to conserve a locality this destroys
a scale invariance. But, in some cases, a scale invariance is more
important physical property than a locality. For this reason one needs to
consider a scale invariant  NSM in other dimensions. Such models can
appear in the local scale-invariant systems with massless fluctuating fields,
which induce an effective action for remaining fields. Usually, the gradient
expansion method is used for obtaining of local effective action.
However, in general, this method does not conserve a scale invariance
of the initial systems. An exact effective action of this systems must be
scale invariant and nonlocal. The last property is connected with
long-range character of interactions induced by massless fluctuating fields.
The well known example of such
interaction in 3D space is the van der Waals interaction. Another
important example is a $1/r^2$ potential in one-dimensional systems \ci{4,5,6}.
The corresponding NSM has many properties common with that of
2D  models \ci{7} and some applications
to various physical systems \ci{8}.

The TE of the scale-invariant NSM also have the interesting properties.
For example, the TE with logarithmically divergent
energy induce in low-dimensional systems with $D \le 2$ the
topological phase transitions (TPT),
which change drastically  the correlations in these systems [9,10,11].
An existence of such TE depends on the nontriviality of
the  homotopic group $\pi_{D-1}({\cal M}).$
\sms

\und {\bf Question} :

\sms

Can such TE and TPT exist in NSM with $D>2?$

\sms

Unfortunately, the TE with logarithmic energies do not appear
in usual higher dimensional theories.

The main efforts in higher dimensional systems  were devoted
to the discovery
of the TE with {\it finite energy} [12,13]. All such
excitations give finite contribution to the partition function,
but cannot induce a PT similar to the TPT, since the latter is induced by
TE with {\it logarithmically divergent energy}.

Recently it was shown that the TE with logarithmic energy can exist
in 3D conformal (or the van der Waals) NSM \ci{14}.
In this talk we introduce the conformal D-dimensional NSM and
show that  they can have different TE, including ones  with logarithmical
energies. An existence of last excitations
is intimately related  with a scale and conformal symmetries of the models.

\bs

\und{\bf II. D-DIMENSIONAL CONFORMAL $\sigma$-MODELS.}

\bs

A condition of existence of
 TE with discrete topological charges and logarithmic
energy puts over on NSM the following properties :

1) their homotopical group
$\pi_{D-1}({\cal M})$ must be nontrivial abelian discrete,

2) a scale invariance at classical level.

The first property
permits some ambiguity in a dimension and a form of ${\cal M}$,
while the second one defines a form of the NSM action ${\cal S}$
almost uniquely in arbitrary dimensions.

A general expression for action ${\cal S}$ of $D$-dimensional generalized
NSM,
admitting nonlocal ones, can be represented in the next form
$$
{\cal S} = \fr{1}{2\alpha} \int d^D x d^D x' \psi^a(x)
\boxtimes_{ab}^{(D)}(\psi|x,x') \psi^b(x'),\quad a,b=1,...,n
\en(1)
$$
where $\psi \in {\cal M},$ and $n$ is its dimension.
The form of the kernel $\boxtimes$ depends
on model. If the structures of the internal and physical spaces do not depend
on each other and the latter space is homogeneous, then $\boxtimes$ can
be decomposed
$$
\boxtimes^{(D)}_{ab}(\psi|x,x') = g_{ab}(\psi(x),\psi(x')) \Box_D (x-x'),
\en(2)
$$
where $g_{ab}$ is some two-point metric function on ${\cal M}.$
For local models an expression for $\boxtimes$ can be defined in
terms of manifold ${\cal M}$ only
$$
\boxtimes(\psi|x,x') = g_{ab}(\psi)\boxtimes \delta(x-x')
\eqno(3)
$$
If the manifold
${\cal M}$ can be embedded in Euclidean vector space ${\mbb{R}}^{N(n)}$
with dimension $N(n),$ depending on $n,$ then one can use instead of
$g_{ab}(\psi,\psi')$ an usual Euclidean metric
$$g_{ab} = \delta_{ab}, \quad a,b=1,...,N.
\en(4)
$$
and the  constraints defining ${\cal M}.$
One must consider the manifolds with a discrete
abelian homotopic group $\pi_{D-1}({\cal M}).$
The spheres $S^{D-1}$ are the simplest among them with
$\pi_{D-1}({\cal M})= \mbb {Z}.$
Then  $N(D-1)= D$, and
$$\psi^a = n^a, \quad a = 1,...,D, \quad ({\bf n})^2 = 1.
\en(5)
$$
where ${\bf n}(x)$ is a field of unit vectors in internal space ${\cal R}^D.$
Since $N = D$  this internal space can be identified
with a physical space.
Below we confine ourselves by the simplest manifolds, the spheres.
A possible generalizations will be shortly discussed in Section Y.

It is more convenient to write ${\cal S}$ and $\Box_D$ in the momentum space
$$
{\cal S} = \fr{1}{2\alpha} \int d^D x d^D x' ({\bf n}(x){\bf n}(x'))
\Box_D(x-x') =
$$
$$
\fr{1}{2\alpha} \int \fr{d^D k}{(2\pi)^D} |n(k)|^2 \Box_D(k)
\en(6)
$$
For asymptotical scale invariance at large scales the kernel $\Box_D$ must
have  the next behaviour at small $k$
$$
\Box_D (k)\simeq |k|^D (1 + a_1 (ka) + ...) = |k|^D f(ka)
\eqno(7)
$$
where $a$ is a UV cut-off parameter, $f(ka)$ is some regularizing function
with the next asymptotics
$$
f(ka) = 1 + a_1 ka + ..., \quad ka \to 0,
\quad f(ka) \to 0, \quad ka \to \infty.
\en(8)
$$
The kernel $\Box_D$ generalizes the usual  Laplace
kernel of two-dimensional $\sigma$-model
$$
\Box_2(k)\equiv \Box(k) = - (\partial)^2(k) = k^2
\eqno(9)
$$
For this reason $\Box_D$ can be considered as a $|\partial|^D$ kernel.
From (9) it follows that in even dimensional spaces $\mbb {R}^{2s}$
the kernel $\Box_{2s}$ is a local one
$$
\Box_{2s} = (-1)^s ((\partial)^2)^s.
\en(10)
$$
 $\Box_D$ is always a nonlocal one in odd dimensions $D=2s+1$
with a large-scale asymptotics
$$
\Box_D(x)|_{x \gg a} \simeq  A_D/|x|^{2D}, \quad
A_D = \fr{2^D \Gamma(D)}{\pi^{D/2}\Gamma(-\fr{D}{2})}.
\en(11)
$$
Such kernels appear often in physics.
The most known and important
cases correspond to $D=1$ \ci{4,5,6} and to $D=3$ \ci{14}.

In the higher dimensional spaces they can describe an effective,
fluctuation induced, interactions.
The similar nonlocal kernels exist in even-dimensional spaces too
$$
\Box_{2s}(x) \sim 1/x^{4s},
\en(12)
$$
but their Fourier-images contain an additional
logarithmic factor
$$
\Box_{2s}(k) \sim k^{2s}\ln (k/k_0)
\en(13)
$$
where $k_0$ is some UV cut-off parameter, regularizing a kernel (11)
at small scales.
This factor  breaks some important properties of the model,
for simplicity,  we will discuss in even-dimensional spaces
only a local kernels.

A conformal group in the higher-dimensional ($D>2$)
spaces is finite-dimensional \ci{15}. Its main nontrivial transformation
is an inversion transformation
$$
x^i \to x^i/r^2, \quad r = |{\bf x}|.
\en(14)
$$
The conformal invariance of ${\cal S}$ with a kernel (11)
follows from the next transformation properties of the kernel $\Box_D$
under conformal transformation (14)
(the field ${\bf n}(x)$ and a coupling constant
$\alpha$ are dimensionless)
$$
x_i \to  x'_i = x_i/r^2, \quad r \to r' = 1/r, \quad x_i/r = x'_i/r',
$$
$$
d^D x \to d^D x/|{\bf x}|^{2D},\quad
\fr{1}{|{\bf x}_1-{\bf x}_2|^{2D}} \to
\fr{|{\bf x}_1|^{2D} \,|{\bf x}_2|^{2D}}{|{\bf x}_1-{\bf x}_2|^{2D}},
$$
and, consequently,
$$
{\cal S} = \fr{A_D}{2\alpha} \int d^D x_1 d^D x_2 \;
\fr{({\bf n}_1{\bf n}_2)}{|{\bf x}_1-{\bf x}_2|^{2D}}
\en(15)
$$
is invariant and dimensionless.
Thus the action (15) with a kernel (11) can be named a  D-dimensional
{\it conformal} NSM.
Strictly speaking, a conformal invariance takes
place only at large scales, since a kernel (11) needs some regularization
at small distancies, which can break this invariance.

The corresponding Euler - Lagrange equation has a form
$$
\int \Box_D(x-x') {\bf n}(x')d^D x' -
{\bf n}(x) \int ({\bf n}(x){\bf n}(x')) \Box_D(x-x') d^D x' = 0.
\en(16)
$$

Its Green function $G^D(x)$  has the following form
$$
G^D(x) = \Box_D^{-1} = |\partial|^{-D} =
\left.\int \fr{d^D k}{(2\pi)^D}\fr{e^{i({\bf k}{\bf x})}}
{\Box_D(k)} \right.
\en(17)
$$
At large scales $G^D(x)$  has a logarithmic asymptotic behaviour
$$
G^D(x)|_{r \gg a} \simeq
- B_D \ln (r/R), \quad
B_D = \fr{1}{2^{D-1}\pi^{D/2}\Gamma(\fr{D}{2})},
\en(18)
$$
where $R$ is a radius of the space or a size of system.
Thus the conformal kernels $\Box_D$ coincide with kernels of
the field theories
equivalent to the generalized logarithmic gases \ci{16}.

\bs

\und{\bf III. TE WITH LOGARITHMIC ENERGY.}
\bs

Since $\pi_{D-1}(S^{D-1})= \mbb {Z}$, there are the TE with topological charge
$Q \in \mbb {Z}$. The simplest TE with charge $Q=1,$ corresponding to the
identical map of spheres $S^2,$ has the next asymptotic form (a "hedgehog")
$$
n^i(x)_{r \gg a} \simeq  \fr{x^i}{r}
\en(19)
$$
The action ${\cal S}$ of this solution for odd $D$ (with logarithmic
accuracy) is
$$
{\cal S} = \fr{{\cal C}_D}{\alpha} \ln (R/a), \quad
{\cal C}_D
= \fr{2^{D-2}\pi^{\fr{D-2}{2}}(D-1)^2 \Gamma^2(\fr{D-1}{2})}
{\Gamma(\fr{D}{2})}.
\en(20)
$$
The interaction of two different TE with charges
$Q_1$ and $Q_2$ on large distancies
has a form of the Green function $G^D(r)$
$$
E_{12} (r) = Q_1 Q_2 G(r) \simeq  -  Q_1 Q_2 B_D  \ln (r/a)
\en(21)
$$
Due to "no hair-dressing theorem", only the "hedgehog" type solutions
can exist in odd dimensional spaces \ci{15}.
In even dimensional spaces the transverse field configurations are
also possible.
The hedgehog energy for even $D$ is
$$
{\cal S} =  \fr{\pi^{D/2} \Gamma(D+1)}
{2\alpha \Gamma(D/2)} \ln (R/a),
\en(22)
$$
 In the usual local D-dimensional NS-model with action
$$
{\cal S} = \fr{1}{2A} \int d^D x (\partial {\bf n})^2
\en(23)
$$
such TE have the energy
$$
E \simeq \fr{S_{D-1}(D-1)}{2 A(D-2)} (R^{D-2}-a^{D-2}).
\en(24)
$$
For the mixed action ${\cal S}_{mix}$,
containing a sum of
kernels $\Box_d, \quad  d = 2,D,$
the corresponding
equation has again the "hedgehog" solution (19) with total energy
$$
E = \fr{S_{D-1}(D-1)}{2 A(D-2)}(R^{D-2}-a^{D-2}) + \fr{{\cal C}_D}{\alpha} \ln (R/a).
\en(25)
$$
where $S_{D-1} = \fr{2\pi^{\fr{D}{2}}}{\Gamma(\fr{D}{2})}$ is a volume of
the unit $(D-1)$-dimensional sphere.

It means that a logarithmic part of the "hedgehog" energy in mixed models
can be observed at scales
$$
(A/\alpha)^{1/(D-2)} > l > a.
$$
The analogous "anti-hedgehog" solutions
with the same logarithmic
energies exist also.

\bs

\und{\bf IY. OTHER TOPOLOGICAL EXCITATIONS.}
\bs

The TE of instanton type  with finite
energy can also exist in the conformal NSM.
They correspond to the configurations
with trivial boundary condition
$$
{\bf n}(x) \to {\bf n}_0, \quad r \to  \infty,
\en(26)
$$
where ${\bf n}_0$ is some constant unit vector. A necessary condition
for their existence is a nontriviality of group
$\pi_D(S^{D-1}).$ Since
$$\pi_D(S^{D-1}) = \mbb {Z}_2,
\; D>3,$$
it means that the D-dimensional ($D>3$)
conformal NSM on spheres
have the instanton-like TE only with $\mbb {Z}_2$ topological charges
$$
Q \in \pi_D(S^{D-1}) = \mbb {Z}_2 = \mbb {Z} (mod 2) .
$$
In case $D=3$ NSM has the TE with finite energies, the hopfions.
They are characterized by topological charge, coinciding with
the Hopf invariant $H \in \mbb {Z}$
of the corresponding mapping $S^3 \to S^2.$
This invariant is connected  with linking number of two
projected circles $S^1$
$$
\{\gamma_1, \gamma_2 \} =
\fr{1}{4\pi} \oint_{\gamma_1} \oint_{\gamma_2}
\fr{<{\bf r}_{12} \cdot [d{\bf r}_1 d {\bf r}_2]>}
{|r_1-r_2|^3}.
$$
For one winding of one circle around another $H = 1.$
If a mapping projects each circle $q_i \;(i=1,2)$ times then $H= q_1 q_2.$
This additional topological invariant classifies "neutral"
(relatively to group $\pi_2(S^2)$)
configurations in all 3D NS-models defined on sphere $S^2.$

\bs

\und{\bf Y. GENERALIZATIONS.}

\bs

The natural generalizations are the conformal NSM on:

1) simple compact groups $G$,

2) their homogeneous spaces $G/H.$

For $G$ not all  homotopic groups  $\pi_i(G)$ are known. For classical
groups all $\pi_i(G)$ are divided
into stable homotopic groups and nonstable ones. The first ones
correspond to
$$i \le A_G\, n + B_G.$$
Here $n$ is a rank of group $G,$
$A_G, B_G$ are some  constants of order O(1), depending on $G.$
For groups $G = SO(n), U(n), Sp(n)$
$$
A_{SO(n)} = 1,\; B_{SO(n)} = - 2;\quad A_{U(n)} = 2,\; B_{U(n)} = - 1;
$$
$$
A_{Sp(n)} = 4, \; B_{Sp(n)} = 1.$$
The stable homotopic groups do not
depend on rank $n$ and are denoted as $\pi_i(G)$.
They are known for all classical groups and have a property
of the Bott periodicity
$$
\pi_i(G) = \pi_{i+b_G}(G)
$$
where $b_(G)$ is a corresponding period. For classical groups
$G = U,SO,Sp$ the Bott periods have the next values \ci{15}
$$
b_{U} = 2, b_{SO} = 8, b_{Sp} = 4.
$$
If one confines himself only by
free (infinite) homotopic groups (without finite parts or torsion) then
the nontrivial stable groups
$\pi(G)$ correspond to the so-called characteristic classes of groups $G$
\ci{15}
$$
\pi_k(G) \ne 0, \quad k = 2 k_i(G) - 1, \quad 1 \le i \le n(G)
\en(37)
$$
where $k_i(G)$ are the Weyl indices of group $G.$
It follows from (38) that
only odd homotopic groups can be nontrivial. All Weyl indices, the degree
of the Weyl invariant polynomials on the maximal abelian Cartan subalgebra,
are known \ci{17}:
$$
k_i(U(n)) = 1, 2,...,n;
$$
$$
k_i(SO(2n+1)) = 2, 4,..., 2n  = k_i(Sp(n));
$$
$$
k_i(SO(2n)) = 2, 4,..., 2n-2; \quad
$$
$$
\mbox {(2n-2 appears twice for even n and once
for odd n)};
$$
There is a simplified form of
the Bott periodicity for infinite homotopic groups of classical groups:
$$
\pi_i(G) = \pi_{i+4}(G) \quad \mbox{for } G = SO, Sp,
$$
$$
\pi_i(U) = \pi_{i+2}(U).
$$
The nonstable groups are finite and are known only in some partial cases.

The TE with logarithmic energy and topological charges $Q \in \mbb {Z}$
are possible only for
even dimensional spaces with
$$D = 2k_i(G),$$
and the
instanton-like TE - only for odd dimensional spaces with
$$D = 2k_i(G) - 1.$$

All above topological charges are scalars (i.e. one-component).
In some cases (for example, for topological interpretation of the quantum
numbers) it is very important to have TE with vectorial topological charges
\ci{18}.

\und {For conformal NS-models such TE can exist only in 3D space.}

To obtain them one
needs to consider NS-models on the maximal flag spaces $F_G=G/T_G$
of the simple compact groups $G.$  They have
$$\pi_2(F_G)= \mathbb {L}_v,$$
where $\mathbb {L}_v$ is a  dual root lattice of $G.$
This lattice, in general, is not enough for topological
description of {\bf all} quantum numbers of group $G,$ (among them are
$SU(n)$ groups).
This fact can be connected with a problem of quark confinement \ci{1,18}.
The 3D conformal NS-models on $F_G$
have the TE with a logarithmic energy
and interacting vector
topological charges ${\bf Q} \in \mathbb{L}_v.$
Since $\pi_3(F_G)= \pi_3(G)= \mbb {Z},$
in this case the "neutral" configurations will also have different
topological structures described by group $\pi_3(F_G).$

For higher homotopic groups $\pi_i(F_G) = \pi_i(G),\quad i>3,$ thus
in higher dimensions ($D>3$) only scalar charges are possible
in these models.

I would like to thank the organizers of the ICMP-2000 conference
for the opportunity to present this talk and support.

This work was supported also by RFFI grant 00-15-96579.
\end{document}